\documentclass[3p]{elsarticle}

\usepackage{booktabs}

\title{Finite temperature dynamics of anionic water trimers}
\author[kcl]{J. M. Finn}
\author[kcl]{F. Baletto\corref{cor1}}
\address[kcl]{Physics department, King's College London, Strand, London, UK, WC2R 2LS}
\cortext[cor1]{Corresponding author: francesca.baletto@kcl.ac.uk}

\begin{document}

\begin{abstract}
\noindent Utilising Car-Parrinello molecular dynamics simulations, the finite temperature behaviour of  water trimers, $\mathrm{(H_2O)_3}$, in both neutral and anionic frameworks, has been investigated. A significant structural change in the anionic structure has been observed at temperatures above 100~K where a chain geometry has formed and stabilised entropically. On the other hand, neutral trimers have remained in their ring structure, as predicted theoretically at low temperatures, for long time periods.   
\end{abstract}

\begin{keyword}
Water trimer \sep Excess electron \sep Finite temperature
\end{keyword}

\maketitle

\section{Introduction}
Water clusters have a vast role within various areas of science, ranging from radiation chemistry \cite{Garrett:2005bw} and molecular biological processes \cite{Garczarek:2006go, Michalarias:2002hk} to the suggestion they play a non negligible role in atmospheric chemistry \cite{Headrick:2001wpa, Lu:2009ga}. The hydrated electron, usually denoted as $e^-_{aq}$, in water systems represents an intriguing and important example of electron solvation, since its discovery in bulk water, where the hydrogen-network (H-network) of six water molecules form a hydration shell with an octahedral symmetry \cite{Hart:1962wl, SCHLICK:1976ug}. Recent experimental and theoretical works have demonstrated that a dynamical rearrangement of the surface free OH-groups occurs both at the ice-vacuum and water/air interfaces to localise an excess electron (EE) \cite{Baletto:2005bj, Bovensiepen:2009kj,Madarasz:2007kp,Sagar:2010, Herbert:2011, Abel:2012}. 

In the last two decades, thanks to the advances in cluster science, significant research has been undertaken into the description of electron hydration in water nanoparticles and how to extrapolate this information to the macroscopic level. Specific challenges include the minimum number of water molecules and geometric structure required to bind an excess electron, which are found experimentally through electron injection time of flight spectroscopy and electron photodetachment spectra \cite{Verlet:2005kq, Ehrler:2009io}. It has been shown that the vertical detachment energy (VDE), which is the vertical binding energy of the excess electron evaluated at the anionic cluster morphology, is related to the localisation/delocalisation of the extra electron, although it does not provide any information regarding the ``position'' of the excess electron with respect to the hydrogen network, i.e. surface or cavity bound. 
Besides several electronic structure calculations devoted to find the best geometrical isomer for anionic water clusters, to estimate their VDE and to determine whether the extra electron is surface or internally bound \cite{Herbert:2011,Tachikawa:2006,Yagi:2008jg,Marsalek:2012}, the current understanding of how subtle thermal fluctuations of the aqueous solution influences the diffusive properties of the EE is still poor, due to the variety of structural motifs that a water cluster has access to during its dynamics.
Recently, through {\em ab-initio} Born-Oppenheimer  molecular dynamics simulations it has been predicted that even on medium size water clusters, the EE is preferentially surface solvated \cite{Frigato:2008hn}. Moreover, free jet ion source experiments have shown that the VDE depends on the different carrier gas used, and thus on the cluster temperature, allowing the formation of different morphologies \cite{Young:2012,Barnett:2011}. 
It is of the utmost importance to have an atomistic description of the finite temperature dynamics of the H-network in order to give insights into the electron transfer processes \cite{Tay:2009bp,Marsalek:2012}. 

In this manuscript for the first time, the linearisation process of the anionic water trimer has been observed through finite temperature Car-Parrinello molecular dynamics.  
Although a linearisation mechanism has previously been explored both experimentally \cite{Liu:1994}, with an estimation of the attachment energy of 142 $\pm$ 7 meV \cite{Herbert:2011,Arnold:1993}, and theoretically \cite{vanderAvoird:2008fo, Taketsugu:2002kc}, to our knowledge this is the first time it has been observed to be a thermally activated process. Here a detailed analysis of the dynamic trajectories has shown that the formation and the stabilisation of a linear chain geometry for an anionic water trimer occurs in a few picoseconds and the linear shape remains stable for at least 15~ps. The structural transformation involves two main steps: the first leads to the loss of a hydrogen bond and the formation of a double acceptor (AA), referring to a single molecule which coordinates both its hydrogen atoms to the extra charge, at the end of the chain. Which is stabilised by a twist of the OH bonds at the other end of the clusters in order to maximise the cluster total dipole moment and to form an effective co-operation among the individual single water molecule dipole moments allowing the additional electron to become bound to the AA.
Our findings could explain the low intensity peak corresponding to the water trimer clusters in the infrared spectroscopic measurements \cite{Ayotte:1999uu}, and are in agreement with other theoretical calculations based on higher level quantum chemical methods, such as second-order M\o ller-Plesset perturbation theory (MP2) \cite{Tachikawa:2006, Hammer:2005iw}. 
Since the water trimer has been previously chosen as preliminary model to describe the H-dynamics in bulk water \cite{Keutsch:2001uf},  a full description of the H-network during this process can be considered as a new and effective electron transfer mechanism whenever the solvent could be described as consisting of small water clusters, and thus has significant relevance to  atmospheric reaction pathways as well as numerous biological systems. 

\section{Computational Methods}

To study the finite temperature behaviour of the water trimer, in both the neutral and charged frameworks, at atmospherically relevant temperatures Car-Parrinello molecular dynamics simulations  \cite{CAR:1985wt} have been performed, using the open source Quantum-Espresso package \cite{Giannozzi:2009hx}. The electrons were attributed a fictitious mass of 400 a.u. and the Brillouin zone sampling was restricted to the $\Gamma$-point only.  A timestep of 1 fs (4 a.u.) was used to integrate the equations of motion. The valence electron-nuclei interactions were described by Troullier-Martins pseudopotentials \cite{TROULLIER:1991wi}. Becke-Lee-Yang-Parr (BLYP) exchange-correlation functional \cite{BECKE:1988tx} has been considered to describe the valence electron-nuclei interactions with a plane wave energy cut-off of 100 Ryd, to enable the system to be computationally tractable and the energy convergence threshold is less than 0.1 mHa.  Periodic boundary conditions have been applied to a cubic box of length 40 bohr, and a multiple vibrational Nos\'{e}-Hoover thermostat \cite{NOSE:1984wa}, characterised by two frequencies of 8 THz and 20 THz to thermalise both the slow oxygen-oxygen and the fast oxygen-hydrogen motion, respectively, has been used. The system has been heated gently to around 50-80K and allowed to equilibrate before being heated again to 120K. The whole thermalisation time was at least 5~ps and the reported data have been accumulated over 15~ps for both the neutral and charged water trimers, with over 100~ps obtained for the neutral trajectory and 25~ps for the higher temperature dynamics. 

The dynamical analysis of the charged system has been performed within the Mauri self-interaction correction scheme (MSIC)  \cite{dAvezac:2005iy}, which has been shown to give good results for the description of water systems \cite{Baletto:2005bj,Bovensiepen:2009kj, Bhattacharya:2010vz}. This framework works in the local spin density approximation where the paired spin up and spin down wave-functions are forced to be the same and the energies associated with the Hartree potential, E$_H$, and the exchange and correlation functional, E$_{xc}$, are corrected such that;

\begin{eqnarray}
\nonumber
E_{M,H}= E_H[ n_{\uparrow}+ n_{\downarrow}]-\epsilon E_{H}[ n_{\uparrow}-n_{\downarrow} ] \\
E_{M,xc} = (1 - \alpha) E_{xc}[n_{\uparrow},n_{\downarrow}] + \alpha E_{xc}[n_{\downarrow},n_{\downarrow}],
\end{eqnarray}

\noindent where $ n_{\uparrow}= n_{\uparrow}(\mathbf{r})$ and $n_{\downarrow}=n_{\downarrow}(\mathbf{r})$ are the charge densities for spin up and spin down, respectively. We have used a modified version of the original MSIC scheme whereby the self-interaction is only partially screened, imposing that the two parameters $\alpha$ and $\epsilon$, introduced by Sprik and coworkers \cite{Vandevondele:2005hh}, are both less than the unit.
The choice of these two parameters follows the requirement to reproduce the electron attachment energy of a water dimer of $26-51$ meV \cite{Bouteiller:1998wi}. Using $\alpha=0.85$ and $\epsilon=1.00$, a corrected value of $49$ meV has been calculated against an enormous binding energy of $911$ meV using an uncorrected local spin density.
Due to the imposed periodic boundary condition, a Madelung correction \cite{MAKOV:1995vk,Dabo:2008eg} has to be added in order to remove the spurious Coulomb interaction among periodic images. This contribution is proportional to $(1-\epsilon)$ because the interaction of the singular occupied molecular orbital with itself has been already partially removed when working in the Mauri scheme.

\section{Results}

The water trimer presents different isomers which can be divided in two large classes: cyclic and linear. 
A structure is labelled as ring or cyclic when three hydrogen bonds with length between 1.2 and 2.7 \AA\ are formed. To be classified as linear there are two requirements that have to be satisfied simultaneously: an inter-oxygen distance R$_{OO}$ larger than 3.8 \AA; and a obtuse internal angle, $\Theta_{OOO}$. 
Among ring clusters a further characterization has been carried out throughout the description of the geometrical position of the dangling O􏰅H moieties, accordingly to their direction perpendicular to the oxygen plane, where up ($u$), down ($d$) are arbitrarily defined as opposite to each other, and planar ($p$) means that the OH lies on the oxygen plane. The planar position is where the free OH group lies in a position $\pm 0.05$~\AA ~from the oxygen plane. Any groups greater or smaller than this are labelled as up and down, respectively. 

This classification is the basis of the proposed Isomer Population Analysis (IPA) which allows the determination of the cluster morphology, including linear shapes, at each frame in the dynamics in order to calculate the frequency of each visitation.

For the neutral case, the most stable structure has been calculated to exist as a cyclical hydrogen network, where two of the dangling OH groups point in the same direction with respect to the oxygen plane, and the other in the opposite direction ($uud$), in agreement with other high level calculations \cite{Keutsch:2003er, Salmi:2009bw}. The next favourable structure is the $uuu$ where all the free OH point in the same direction, and the final ring considered is the $upd$ which has been found to be the transition state between two degenerate $uud$ isomers. Each of the six equivalent $uud$ isomers are, indeed, easily accessible through an external hydrogen flip through the oxygen plane. The activation energy barrier between two neighbouring $uud$ neutral water trimer is of the order of 38 meV, calculated by means of nudged elastic band \cite{Henkelman:2000tl}. On the other hand, the transition from the $uud$ to the $uuu$ structure presents an activation energy barrier of 68 meV.

\begin{figure}[ht]
 \begin{center}
  \includegraphics[viewport = 0 0 1024 768, width=0.5\textwidth]{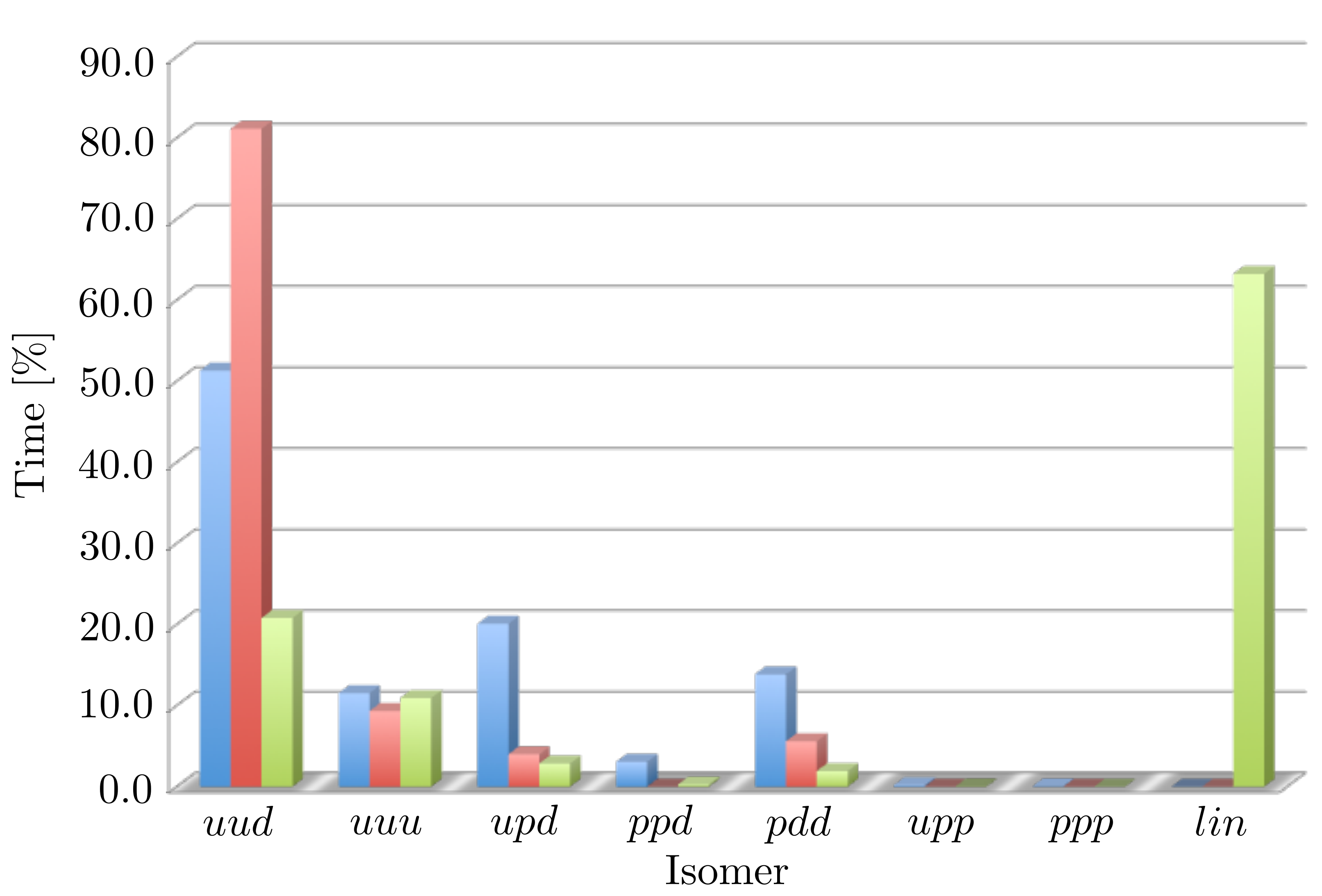}
  \caption{Comparison of the IPA for both the neutral (blue) and anionic 80K (red) and 120K (green) finite temperature dynamics of the water trimer cluster, clearly showing that for the anionic cluster at 120K the linear isomer is chosen preferentially, whereas for the neutral and anionic 80K dynamics the ring isomers are the only structures found. The time is expressed as a percentage of the total simulation time and is unit-less.}
  \label{fig:IPA}
  \end{center}
\end{figure}

A neutral cluster, starting from a $uud$ isomer, has been allowed to evolve at 120~K  for 100~ps and the analysis of its dynamics is reported in blue boxes in Figure \ref{fig:IPA}. No structural changes from the cyclical isomers have been observed, the $uud$ remains stable for around 50\% of the total simulation time, with a only a short time interval (10 \%) spent in the less energetically favourable $uuu$ geometry. A frequent hydrogen motion has been detected and the transition between two equivalent $uud$ rings has been observed in agreement with the nudged elastic band calculations of hydrogen flipping. It is worth stressing that at no point in time is the linear structure observed, as the energy loss due to the breaking of a hydrogen bond is much too high ($\sim$ 0.2 eV).

Two anionic simulations have been performed, at temperatures above and below 100~K. For the lower temperature the simulation has been started from the $uuu$ isomer and has been allowed to evolve for 10~ps, where like the neutral dynamics, no structural transition have been detected from the ring series of isomers with the $uud$ shape found more frequently than the $uuu$, although the latter has a higher total dipole moment. Over the total time period the internal angles are not observed to change further than 60$^\circ$ $\pm$ 10$^\circ$, and the ionic temperature remains relatively stable. The IPA shows that for over 80\% of the simulation time the $uud$ isomer is found, with a further 9\% in the $uuu$, as labelled by red boxes in Figure \ref{fig:IPA}.

The second anionic molecular dynamics has been performed at a higher temperature, again starting from the uuu isomer, and heated using a series of thermostats to around 120~K and then left free to evolve within the Car-Parrinello scheme. The total thermalisation time including the thermostats was around 8~ps, which were discounted from the analysis. Within 2~ps after the equilibration one of the hydrogen bonds broke causing an increase in the maximum internal oxygen angle, indicating the `opening' into the linear isomer. 
The chain isomer is observed to last for around 15~ps, when the ring series of isomers are reformed and remain stable for an additional 2~ps. IPA analysis of the trajectory shows that the linear isomer is dominant for around 62\% of the simulation time, with a secondary isomer of the $uud$ structure, as depicted by green boxes in Figure \ref{fig:IPA}.

\begin{figure}[t]
 \begin{center}
  \includegraphics[viewport = 0 0 533 507, width=0.5\textwidth]{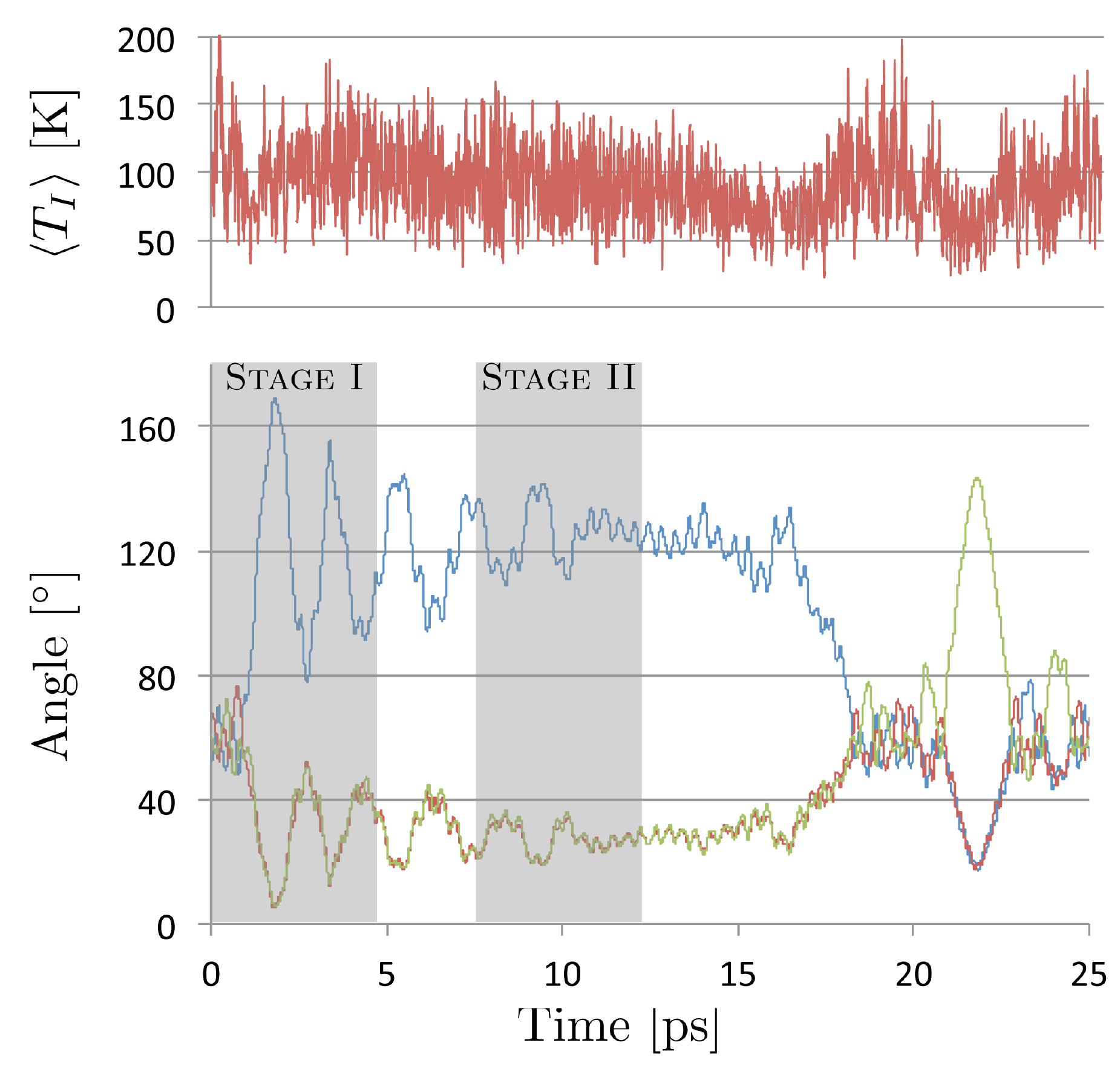}
  \caption{Evolution of the internal oxygen angles within the water trimer at temperatures around 100~K. Top panel shows the temperature evolution (K) and bottom panel shows the two internal oxygen angles ($^\circ$). Shaded regions indicate the two stages of the dynamics, as described in the text.}
  \label{fig:oangle}
  \end{center}
\end{figure}

Two distinct mechanisms occurred during the total dynamics which allowed the water trimer to temporarily remain stabilised in this maximal dipole form. The first, highlighted by the shadow region in Figure \ref{fig:oangle} labelled with \textsc{Stage I}, is the breaking of the hydrogen bond, thanks to a rotation of a free OH group which causes torsional stress on the bond itself. This allows the water molecule, previously donating its hydrogen, to rotate away from the acceptor molecule, and freeing both OH groups. However, the cluster is still not stable and large oscillations of the internal angle cause the isomer to attempt to recombine into the ring structure on the very short time scale, as reported in the bottom panel of Figure \ref{fig:oangle}.  The second mechanism, grey region on the bottom panel of Figure \ref{fig:oangle} titled \textsc{Stage II}, describes a bifurcation flip, whereby the water molecule is seen to rotate through a large angle perpendicular to the oxygen plane, initially breaking the hydrogen bond being donated to the middle molecule in the linear chain. As the rotation changes the hydrogen network and the bond breaks, the second hydrogen moves into the same position in order to form again the chain structure, thus the free OH groups are swapped. 

\begin{figure}[t]
 \begin{center}
  \includegraphics[viewport = 0 0 1252 566, width=0.5\textwidth]{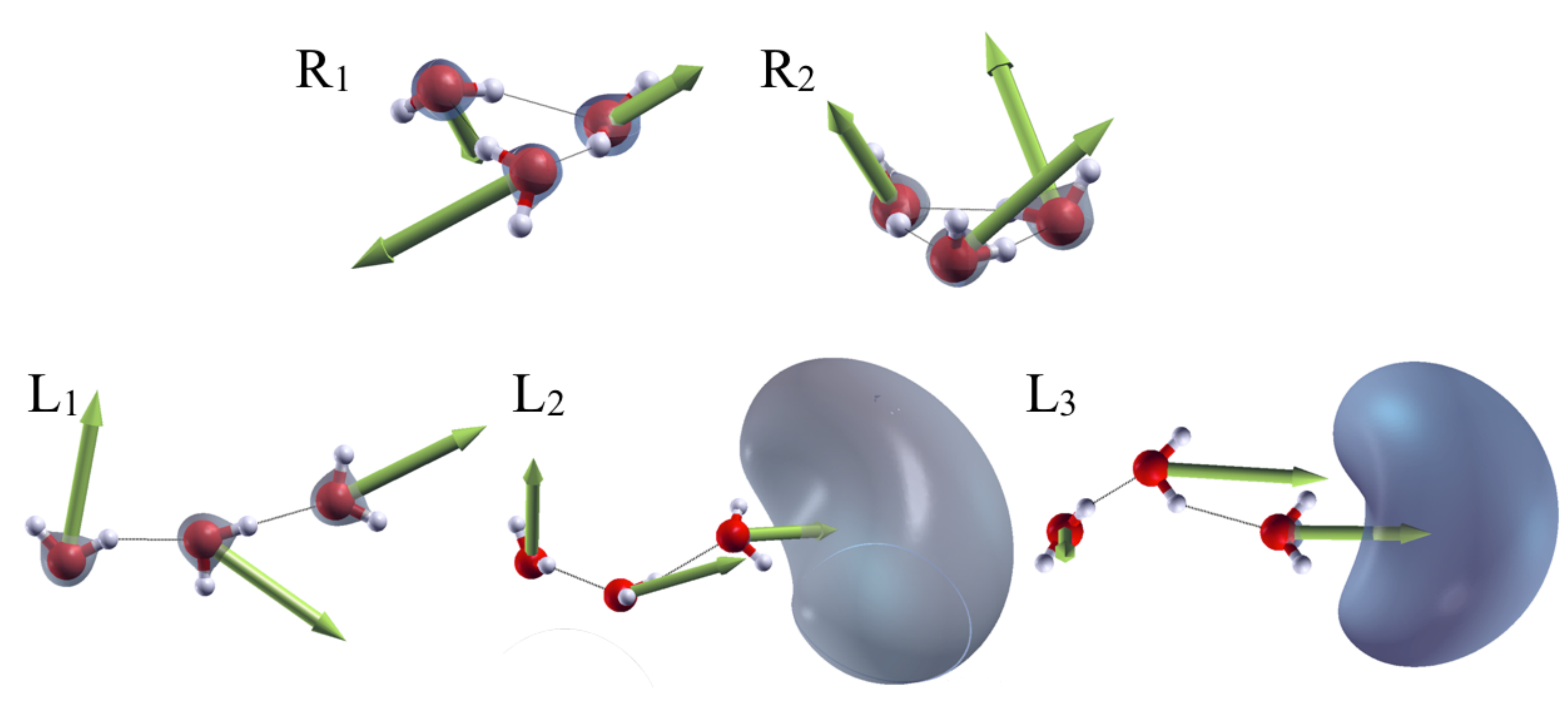}
  \caption{Isosurfaces depicting 70\% of the total charge of the excess electron in differing water trimer isomers. In addition green arrows show the individual dipole moments for each water molecule: notice the uncooperative dipole moments in the $uuu$ structure, and the cooperative binding allowing for the localisation in the linear isomers.}
  \label{fig:ISO}
  \end{center}
\end{figure}

Static analysis of few structures, taken from the dynamical trajectory, gives some indication of the driving force behind the linearization process observed at the higher temperature. These structures are depicted in Figure \ref{fig:ISO}, the cyclical $uud$ (R$_1$) and $uuu$ (R$_2$) motifs in the top row, and the linear shape where L$_1$ was constructed by hand with an internal angle of 180$^\circ$ and L$_2$ and L$_3$ are configurations from the higher temperature dynamics at the end of STAGE I and STAGE II, respectively. Using the partial MSIC correction of $\alpha = 0.85$ and $\epsilon = 1$ different snapshots have been analysed and VAE calculated (table \ref{tab:VAE}). Corresponding calculations using the B3LYP xc-functional confirm the results that the $uud$ is the least bound structure, and L$_3$ is the strongest. All of these results are comparable with literature values, using both MP2 and CI approaches, with strengths of -60 meV and -243 meV respectively \cite{Iwata:2005dw, Tsurusawa:1998uu}. 

The spin density difference between the spin-up and spin-down charge densities population gives an indication of the position of the additional charge: isosurfaces have been plotted in Figure \ref{fig:ISO} depicting 70\% of the total charge. In agreement with the VAE results both the linear structures taken from the dynamics show spatial localisation of the additional charge, at the double acceptor end of the chain. Moreover, it was observed that not only was the localisation and formation of the anionic structure due to the increase in dipole moment of the total cluster but also due to the directionality of the individual dipole moments of each water molecule. The dipole moment of the $uuu$ isomer, which is above the critical amount required to bind the additional charge, however, was crucially not found to attach the additional charge, but a structure which had a co-operative dipole moment formation such as L$_2$ which has a smaller dipole does attach, with only a small difference in the attachment energy. Additionally, once the second stage of the dynamics had occurred, and the final linear isomer was formed (L$_3$) the additional charge became bound to the structure with an attachment energy of -205.5 meV, and spatially was observed to form a bean shaped cloud near to the double acceptor at the end of the chain. Thus this co-operative effect of the individual dipole moments of each water monomer in the cluster is found to have a strong impact on the formation and localisation of the additional charge. 

To support the idea that the dynamics are driven through a kinetic attachment the vibrational density of states (VDOS) has been calculated 
\begin{equation}
VDOS(\omega) = \left( \frac{1}{\sqrt{2\pi}}\int_{-\infty}^{\infty}e^{-i\omega t}Z(t) \mathrm{d}t \right) ^2
\end{equation}
\noindent where $Z(t)$ is the velocity autocorrelation function
\begin{equation}
Z(t) = \frac{\langle \mathbf{v}(0) \cdot \mathbf{v}(t) \rangle}{\langle \mathbf{v}(0) \cdot \mathbf{v}(0) \rangle}.
\end{equation}
Using the VDOS the entropic contribution for each of the finite temperature trajectories can be calculated \cite{Ndongmouo:2007fj, deSousa:2006vi}:

\begin{equation}
T\Delta S = 3k_B\int^{\infty}_0 VDOS(\omega) f(\omega) d\omega,
\end{equation}

\noindent being $f(\omega) = \left[  \frac{\hbar\omega}{2k_BT} \coth( \frac{\hbar\omega}{2k_BT}) - \log(2\sinh( \frac{\hbar\omega}{2k_BT}))\right]$ where and $k_B$ is the Boltzmann constant and $T$ is the temperature. The change in entropy between the uud and linear structures within the anionic framework is found to be -0.025 eV/K, showing that the linear form is entropically stable at higher temperatures, which confirms that there is a kinetically driven attachment of the additional charge. 

\begin{figure}[t]
 \begin{center}
\hspace{-3cm}  \includegraphics[viewport = 0 0 512 380, scale=0.33]{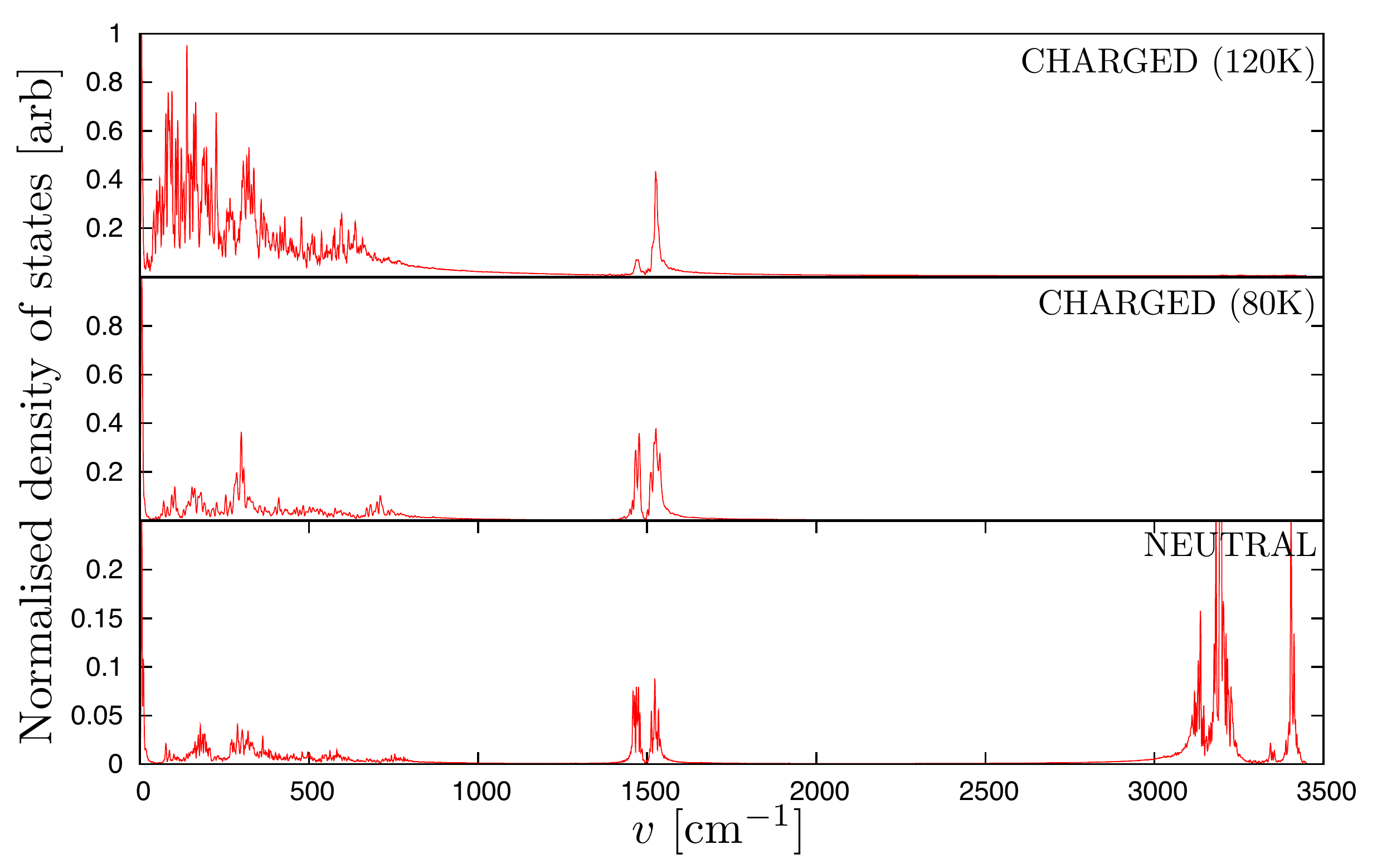}
  \caption{Vibrational density of states for the three water trimer dynamics, neutral and anionic. Frequency is in cm$^{-1}$ and normalised density of states is in arbitrary units. Neutral dynamics at 120~K.}
  \label{fig:VDOS}
  \end{center}
\end{figure}

The VDOS for each trajectory has been plotted in Figure \ref{fig:VDOS} where finite temperature dynamics in the neutral framework agree well with available calculations by Kang \textit{et al.} \cite{Kang:2010wn}, with each of the three bands being expressed. The first band, 0-1000~cm$^{-1}$ shows more peaks in the anionic dynamics corresponding to vibrations in the cluster during the opening and subsequent stabilisation of the chain isomer. The attachment of the excess electron is suggested by the distinctive peak at 250 cm$^{-1}$ seen during both anionic dynamics. The formation of a double-acceptor single molecule can be attributed to the different shape of the vibrational bend peak around 1500 cm$^{-1}$, in agreement with experimental observation \cite{Roscioli:2006wu}. 
However, for both anionic trajectories the third band, between 3000-4000 cm$^{-1}$ corresponding to intra-molecular OH stretching, is under expressed.

\section{Discussion}

Car-Parrinello molecular dynamics simulations of both neutral and charged water trimers have been performed at temperatures relevant to atmospheric processes. Our simulations have shown a direct correlation between the directionality of the individual dipole moments of each water monomer and attachment and spatial localisation of the excess electron. Furthermore, it has been observed that when the anionic system was at a temperature above 100~K the linear formation of the cluster is dominant for around 62\% of the total simulation time, with a temporal stability of around 15~ps.

The fast transition from the ring to linear isomers at 120~K indicates that not only is the energetic stability important during the formation of a water cluster anion, but also that the total dipole, enabling the binding and localisation of the additional charge plays an important role. 

The dipole decomposition has allowed the characterisation of the linear isomers into two types, the first relates to the period of time prior to the bifurcation flip, where the cluster attempts to form co-operative dipole moments in order to maximise the dipole along a vector towards the charge cloud. The second type occurs subsequent to the flip, where the dipole moments are able to align towards the additional charge and therefore, localise the excess electron, co-operatively binding through multiple dipole moments pointing towards the charge. These two effects appear to be behind the driving force of the dynamics.

Thus, the co-operative dipole binding could have significant implications for larger water clusters, and could potentially be behind the general method for binding an additional charge. Secondly an entropic effect has been observed allowing the formation of the linear anion, and the attachment and localisation of the additional charge when the temperature is in excess of 100K. This implies that modelling the finite temperature behaviour of the attachment of excess electrons is important in order to understand the binding motifs, and have greater applicability to the atmospheric anionic formation.  

\section{Acknowledgments}

The authors are grateful to the UK research council EPSRC for its financial support. JF appreciate the HPC2-Europa and Cineca facilities for their financial and computational support. Authors would like to thank the King's College London system manager, Dr. A. Comisso, for the management of local computational facilities.

\begin{table}
\centering
\begin{tabular}{lccc}
\toprule
			&	& \multicolumn{2}{c}{$E_{\mathrm{VAE}}$ [meV]} \\
		\cmidrule{3-4}
 Structure		& $\mu$ [D] & SIC-CP[0.85,1]	& B3LYP	\\ 				
\midrule
\multicolumn{4}{c}{Ring motif} \\
uud	(R$_1$)		&	1.24	 		& -63.1		&  66.2		\\ 		
uuu (R$_2$)			&	4.05			& -86.5		& -0.04   		\\	 	
upd			&	0.98			& -62.0		& 98.6     		\\		
\hline
\multicolumn{4}{c}{Linear motif} \\ 
L$_1$		&	3.94			& -107.8 		& -70.0 	       	\\		
L$_2	$		&	3.70			& -86.3		& -42.5		\\ 		
L$_3$		&	4.49			& -205.5		& -97.4		\\		
\bottomrule
\end{tabular}
\caption{Vertical attachment energy  ($E_{VAE}$ in meV) and total dipole moments ($\mu$ in Debye) for the considered structures reported in Figure \ref{fig:ISO}. Results are compared with the B3LYP functional. A negative value in the VAE refers to a bound EE. Experimental reference value is a bound state at $142 \pm 7$ meV, taken from Ref. \cite{Herbert:2011}.}
\label{tab:VAE}
\end{table}

%


\end{document}